\documentstyle[12pt,aaspp]{article}
\input epsf
\def\l*{$L_*$\/}
\def\etal{{\it et al. }}
\def\vcirc{{v_{circ}}}
\def\kms{\rm ~km~s^{-1}}
\def\msun{M_\odot}
\def\vstar{{(v/\sigma)^*}}

\def\l*{$L_{*}$}

\def\gsim{ \lower .75ex \hbox{$\sim$} \llap{\raise .27ex \hbox{$>$}} }
\def\lsim{ \lower .75ex \hbox{$\sim$} \llap{\raise .27ex \hbox{$<$}} }
\def\pp{\noindent\parshape 2 0truecm 16.0truecm 2.0truecm 15truecm}

\def\spose#1{\hbox to 0pt{#1\hss}}
\def\simlt{\mathrel{\spose{\lower 3pt\hbox{$\mathchar"218$}}
     \raise 2.0pt\hbox{$\mathchar"13C$}}}
\def\simgt{\mathrel{\spose{\lower 3pt\hbox{$\mathchar"218$}}
'     \raise 2.0pt\hbox{$\mathchar"13E$}}}

\begin{document}

\title{Morphological Transformation from Galaxy Harassment} 

\vskip 0.5truecm

\centerline{\bf Ben Moore $^{\bf 1}$, George Lake$^{\bf 2}$ \& 
Neal Katz$^{\bf 3}$}

{\bf 1.} {Department of Physics, University of Durham, Durham, DH1 3LE, UK}

{\bf 2.} {Astronomy Department, University of Washington, Seattle, USA}

{\bf 3.} {Department of Astronomy, University of Massachusetts, Amherst, USA}

\begin{abstract}

Galaxy morphologies in clusters have undergone a
remarkable transition over the past several billion years.  Distant
clusters at $z \sim 0.4$ are filled with small spiral galaxies, many of
which are disturbed and show evidence of multiple bursts of
star-formation.  This population is absent from nearby clusters where
spheroidals comprise the faint end of the luminosity function. 
Our numerical simulations
follow the evolution of disk galaxies 
in a rich cluster owing
to encounters with brighter galaxies and the cluster's tidal 
field---galaxy harassment. 
After a bursting transient phase, they undergo a complete
morphological transformation from ``disks" to``spheroidals".  We
examine the remnants and find support for our theory in detailed
comparisons of the photometry and kinematics of the spheroidal
galaxies in clusters.  Our model naturally accounts for the
intermediate age stellar population seen in these spheroidals as well as the
trend in dwarf to giant ratio with cluster richness. The final shapes
are typically prolate and are flattened primarily by velocity 
anisotropy. Their mass to light ratios are in the range 3---8 in
good agreement with observations.

\end{abstract}


\section{Introduction}

The origin of the Hubble sequence (Hubble 1936, Sandage 1961) remains
a long standing puzzle in astronomy.  Giant galaxies 
range from ellipticals, slowly-rotating dense spheroids with little
gas, to late--type spirals that are rapidly rotating thin disks of
gas and stars.  In recent years, two distinct classes of ellipticals
have been recognized that are easily separated in plots of nearly any
two of their
properties, such as central surface brightness versus luminosity
(Ferguson and Binggeli 1994, Kormendy 1985).  One class includes the
bright, giant ellipticals and extends to the rare ``dwarf ellipticals"
with high surface brightness, M32 being the prototype.  The second
class consists of low surface brightness spheroidal galaxies that have
luminosities $M_{_B} \gsim -17$, about 3 magnitudes fainter 
than \l*, the
characteristic break in the luminosity function.  (Throughout this
paper, all distance dependent quantities assume a Hubble constant of
$100h \kms Mpc^{-1}$, with $h=1$.)  The dwarf spheroidal galaxies
(dSph) in our Local Group of galaxies with magnitudes in the range $
-8 \gsim M_B \gsim -12$ are often considered to be the low
luminosity extreme of this sequence, but nearly all other known
galaxies in this class reside in clusters (Vader and Sandage 1991).
This second class is referred to as ``dwarf ellipticals'' by Ferguson
and Binggeli (1994) and ``spheroidals'' by Kormendy (1985) who
reserves the designation ``dwarf ellipticals'' for galaxies like M32.
After attempting a clumsy hybrid of labels to stay clear of this
confusion, we reluctantly follow Kormendy.

There is no shortage of theories for the formation of giant galaxies
and the origin of their Hubble sequence.  Hoyle (1945) calculated an
upper limit to the mass that can radiatively cool and proposed that
this imprints a characteristic galaxian mass scale.  Gott (1977) and
Gunn (1982) promoted two epochs of galaxy formation with ellipticals
and bulges forming in an early epoch of Compton cooling and spirals
forming by radiative cooling at a later time.  This model has faded
from popularity because the two formation epochs 
means that morphological types must arise from
fluctuations with vastly different amplitudes, must cool by
different physical processes yet have comparable velocity 
scales and masses.

%***the last sentence needs clarifying

Most theories for the origin of 
the Hubble sequence follow Jeans' (1938) assertion
that angular momentum is the control parameter.  Among the giant 
galaxies, the
Hubble sequence can also be viewed as a mass sequence (Tully, Mould
and Aaronson 1982) or virial velocity sequence (Meisels 1983).  The
larger ellipticals naturally occur in clusters owing to statistical
biasing in the hierarchical gravitational collapse model (Bardeen
\etal 1986), where larger peaks are found to be more strongly clustered
than smaller ones.  Unfortunately, the variation in the angular
momentum of protogalaxies in the hierarchical model is small compared
to the observed variation across the Hubble sequence. Furthermore, 
the angular momentum of halos correlates poorly with 
any other property 
such as local overdensity (Barnes and Efstathiou 1987, Ryden 1988).

In order to match the wide variations in the observed properties
of galaxies we need a mechanism that can increase the
available stretch and correlations in the hierarchical model.
One way to do this is to form elliptical galaxies by
the merger of two spirals (Toomre 1977).  Despite continuous
assertions of its demise (Ostriker 1980, van den Bergh 1982, Carlberg
1986), it seems fairly robust (Lake and Dressler 1986, Lake 1989, Zepf
and Ashman 1993).  However, ellipticals are generally found within 
larger structures with 
collapse times that are shorter than the cooling time
required to make a disk.  Lake \& Carlberg (1988ab) and Katz (1991)
found that the Hubble sequence could also arise if the relative
amplitude of small scale fluctuations increased with mass or binding
energy.  If larger mass fluctuations have greater small scale power,
then their lumps lead to a transfer of angular
momentum from the luminous
material to the dark matter.  An elliptical galaxy easily results from
a lumpy collapse model without disks having already formed.
Similarly, Lake \& Carlberg noted that small scale power must be suppressed in
order for a disk to survive, an effect that was reexamined by Toth and
Ostriker (1992).

Spheroidal formation theories have a shorter history and have focused
on incremental changes to the giant galaxy theory.  The popular model
of Dekel and Silk (1986) combines the notion of lower amplitude peaks
in the hierarchical model with the use of stellar winds or supernovae
to reshape the small galaxies that have low binding energy by
expelling their gas (Larson 1974, Vader 1986).  There are many
problems with this scenario, {\it e.g.} the clustering
properties of dwarfs is opposite to the expectations of 
the Dekel and Silk model (Ferguson and Binggeli 1994). The model
also has the seemingly impossible chore of explaining the general
properties of both rapidly-rotating gas-rich dwarfs and gas free dwarf
spheroidals.
% avoiding conflict with the survival of globular
% clusters.

Recent Hubble Space Telescope (HST) observations (Dressler \etal
1994a) have revealed that the morphologies of galaxies in clusters
have changed dramatically since $z \sim 0.4$.  Over 20 years ago,
Butcher and Oemler (1978, 1984) discovered a large population of
``blue galaxies" in clusters at $z\sim 0.4$.  The HST frames show that
at $z \sim 0.4$, the giant ellipticals are already in place, but the
ubiquitous ``blue galaxies" are distorted spirals that have vanished
from clusters at the present epoch (Dressler \etal 1994a).  The
population difference is greatest at lower luminosities; 90\% of
galaxies fainter than $L_*/5$ in distant clusters are bulgeless
``Sd'' disk systems, whereas 90\% are spheroidals in nearby clusters
such as Virgo or Coma (Sandage \etal 1985, Thompson and Gregory
1993).  Couch \etal (1994) present spectroscopic evidence that the
distorted blue galaxies at $z \sim 0.3$ have undergone multiple burst
events separated by 1--2 Gyr.

In hierarchical clustering models, the influx of field galaxies into
clusters peaks at $z \sim 0.4$ (Kauffmann 1995).  A complete picture
emerges of both the Butcher--Oemler effect 
and the morphological
transformation if we can identify a mechanism that begins when spiral
galaxies enter clusters, causes a few episodes of distortion and
star formation within each galaxy over an interval of 1--2 Gyr and
ultimately turns the dwarf spirals into spheroidals in 4--5 Gyr.

Several possible mechanisms have been proposed for the Butcher--Oemler
effect: mergers (Icke 1985, Miller 1988), gas compression in the
cluster environment (Dressler and Gunn 1983) and tidal compression by
the cluster (Byrd and Valtonen 1990, Valluri 1993).  Nearly all of
these mechanisms can make starbursts, but they all fail other critical
tests (c.f. Valluri and Jog 1993).  Merging is a one time event
in contrast to the spectroscopic evidence of multiple bursts found
by Couch \etal (1994). 
From their HST images, Oemler \etal (1996) conclude that merging is
implausible as the blue galaxy fraction is large and the merging
probability is low.  They also observe disturbed spirals throughout
the cluster, whereas both ram pressure stripping and global tides 
only operate efficiently near the cluster's center.  
Furthermore, Valluri
and Jog (1993) have shown that the observed relationship between HI
deficiency and size is exactly opposite to that expected from ram
pressure or evaporative models. Finally, there is no correlation
between the spiral fraction and X-ray luminosity of clusters.
None of the other models for the Butcher--Oemler effect 
discussed above address the
issue of present day remnants: how are the ubiquitous small
spheroidals created as the distorted spirals disappear?

Recently, we proposed a new mechanism for the Butcher--Oemler effect
in clusters---``galaxy harassment" (Moore \etal 1996).  At speeds of
several thousand kilometers per second, close encounters with bright galaxies 
cause impulsive
gravitational shocks that can severely damage the fragile disks 
of Sc--Sd galaxies.  We will show that these collisions are frequent
enough that 
harrasment occurs throughout the cluster. 

Moore \etal (1996) used numerical simulations to compare 
harassed galaxies to HST frames of
galaxies in clusters at $z \gsim 0.3$.  They stated that the
cumulative effect of such encounters changes a disk galaxy into a
spheroidal galaxy, thus identifying the present-day remnants of the
disturbed blue galaxies and explaining the change in galaxy
morphologies in clusters since $z \sim 0.4$.  In this paper, we
provide detailed comparisons of the harassed remnants of late
type spirals with the photometric and kinematical 
properties of dwarf spheroidal galaxies.

\section{Simulations of galaxy harassment}

We use  N-body simulations with only gravitational forces
as well as gas dynamical simulations using TREESPH (Hernquist and Katz 1989, Katz, Weinberg
and Hernquist 1996) to follow the evolution of ``high
resolution'' model spiral galaxies as they orbit within a 
rich galaxy cluster.

The cluster model is based on Coma.  Its
one dimensional velocity dispersion is 
$\sigma_c=1,000 \kms$ and the
mass within the virial radius (1.5 Mpc) is $7\times
10^{14}M_\odot$.  For an $M/L$ of 250, the total
cluster luminosity within this radius is $2.8\times 10^{12}L_\odot$.
The other galaxies in the cluster are drawn from 
a Schechter (1976) luminosity function parameterized using
$\alpha=-1.25$ and $M_*=-19.7$  
including all galaxies brighter than
$2.8\times 10^8L_\odot$ ($H_o=100{\kms Mpc^{-1}}$ and $\Omega=1$
throughout this paper). 
This produces a  model cluster that has 950
galaxies brighter than the Magellanic clouds, but only 31 brighter
than $L_*$.   

The masses and tidal radii of the other galaxies are determined
by taking an
isothermal model with a dispersion given by the
Faber-Jackson relation and 
tidally limiting it at the galaxy's pericentric distance.
They are then modeled by spheres with a softening length
equal to half of their tidal radius.  Most simulations have
no ``interpenetrating" collisions.  However, any such collision
will be more gentle than a collision with a realistic model
of a galaxy that is more centrally concentrated.  
White and Rees (1978) speculated that the dark halos were stripped
from galaxies within clusters;  we find that the dominant
stripping mechanisms are tides and high speed 
encounters---``galaxy harassment" (Moore, Katz and Lake 1996;  hereafter MKL).  
If galaxies are initially 
tidally limited by the cluster potential, 
bright galaxies retain more
than half of their mass when followed for 5 Gyr.  
For greater self-consistency,  we reduced the mass of each
perturber by 25\%, the average loss over a Hubble time (MKL).  
We left the tidal/softening radius fixed.  In the accompanying
video, the perturbing galaxies are shown as green dots located
at their centers.

The mean ratio of a perturbing
galaxy's apocenter to its pericenter in our cluster model is roughly 6-to-1
(this ratio is even larger in infinitie isothermal spheres with isotropic
velocities);   a galaxy found at a radius of 450 kpc will have a 
mean orbital radius of 400 kpc and a 
typical pericenter ($r_{peri}$) that is slightly greater than 150 kpc.  
We assign galaxies masses of $2.8\times10^{11} (r_{peri}/150{\rm kpc})
(L/L_*)^{3/4} M_\odot$ corresponding to mass-to-light ratios, 
$M/L = 26 h^2  (r_{peri}/150{\rm kpc})(L/L_*)^{-1/4}$.
The luminous parts of elliptical galaxies  have mass-to-light ratios 
of $12  h M_\odot/L_\odot$ (van der Marel 1991), so our 
perturbing galaxies have very modest dark halos. 

The fraction of the cluster's density attached to
galaxies varies with radius from zero at the center to nearly
unity at the virial radius.
It is  $\sim$20\% at the mean orbital radius of our simulated
galaxies.  
The rest of the cluster mass is in a smoothly
distributed background represented by a fixed analytic potential.
Further details of the cluster model 
can be found in MKL.
 
The simulations were originally performed to follow 
the morphological evolution of galaxies in clusters (Moore \etal 1996, 
MKL), so the parameters were chosen to be typical of the disturbed
galaxies seen by HST at redshifts of $\sim 0.3$.
Since we are examining the assertion that the remnants become the  
spheroidals seen in present-day clusters, we must compare the
remnants in our simulations to the  brighter
spheroidals observed in clusters.

We simulated a number of galaxies  
with circular velocities of 
$160 \kms$ and $110 \kms$.  
Using the Fisher--Tully (1981) relation,
the luminosity of a galaxy with a circular velocity, $\vcirc = 160 \kms$
is  $\sim L_*/5$, while the  $\vcirc = 110 \kms$
would correspond to  $\sim L_*/20$.  
The $L_*/5$ model galaxies  have
exponential disks with scalelengths of 
2.5 kpc and
scaleheights of 200 pc.   The $\sim L_*/20$
models had disk masses that were half of the $L_*/5$ models
and scalelengths of 2 kpc, giving
an initial disk surface density about 20\% lower than the larger
models. 
They were all constructed with a Toomre (1964)
``stability'' parameter $Q=1.5$ and  run in isolation for 2 Gyr
before being set into orbit in the cluster.  Gas disks were only
used in the larger galaxies and were started with particles on
cold circular orbits.  

The dark halo of each galaxy is spherical and
isothermal with a core radius of $(v_{circ}/160\kms)^2$ kpc
and is tidally truncated at
the pericenter of the galaxy's orbit within the cluster.  
At 4 disk scalelengths (10 kpc) the initial ratio of 
dark matter to stars to 
gas
is 11:4:1 in the $L_*/5$ models.   At 8 disk scalelengths (20 kpc), 
the ratios are
20:5:1 and the total mass is $\sim
10^{11} (v_{circ}/160\kms)^3 M_\odot$. 
We fixed the mean radius of the orbits at 450 kpc as a representative value
that prevented the dimensionality of the parameter space from getting
out of control.

The gas was treated using smoothed particle hydrodynamics that models
the effects of shocks and cooling assuming using primordial element
abundances.  The simulations with gas have no prescription for star
formation, representing an extreme case.  To explore the opposite
extreme, we also ran simulations without gas to represent the case
where the gas forms stars the first time it is disturbed.  The 15
models that were run are all tabulated in Table 1.  These models were
run in a variety of ways: 1) with and without gas included (the
combined mass of the stellar and gaseous disk is fixed), 2) with and
without the perturbing influence of other cluster galaxies and 3) on
circular and elliptical orbits.

Of the 15 simulations
shown in Table 1,
models 1--10 and 11--15 have  
$\vcirc = 160 \kms$ and  $\vcirc = 110 \kms$ respectively.  
Apart from the initial conditions of the two separate
models, all the quantities are measured after 3--5 billion years of
evolution in the cluster. (The runs with gas were stopped 
when the central gas density became so large that the 
required timestep became prohibitively small.)

We have intentionally made some conservative assumptions 
to ensure that our results are robust.  
For a model galaxy on a fixed orbit, 
the havoc wreaked by harassment depends on the square of the masses
of the largest galaxies encountered.
The most massive galaxies are  giant ellipticals
that are far less prone to harassment owing to their high 
internal densities,
yet we have reduced the masses of all galaxies by the same 
time averaged value of 25\%.  
At a fixed mean orbital radius, 
galaxies on elongated orbits experience greater harassment.  
We follow galaxies that have apo/peri ratios of 2 ({\it i.e.} apocenter
at 600 kpc, pericenter at 300 kpc), whereas the typical value is
$\sim 6$.  
As a result, our model galaxies avoid
extremes of the cluster distribution and start with large dark halo
masses determined by the tidal limit at their atypically large pericenters.  
Both effects serve to underestimate the effects of
harassment.

In MKL we examined the evolution of spherical
systems owing to the individual and combined actions of tides and
collisions with other galaxies.  We noted the transformation of small
disk galaxies into spheroids when we proposed that harassment
causes the Butcher--Oemler
effect (Moore \etal 1996).  Here, we present detailed photometric and
kinematic properties of the harassed remnants and compare to
spheroidals in clusters.

The evolution proceeds in a violent, chaotic fashion that is best
appreciated by watching the accompanying video.  
The first sequence shows the orbiting galaxy
and the entire simulated cluster.   It may appear that global cluster tides
are stripping most of the material from the model, but
it is the collisional shocks from other galaxies that 
drives the bulk of the dark
matter over the tidal radius
(a model galaxy on the same orbit but with no perturbers
looses a great deal less mass).  The stripped material continues to follow the
orbital path of the galaxy and is seen as the prominent 
tidal tails. 
The third dynamical
sequence (the stellar distribution framed in an 18 kpc box) shows the
transformation from disk to spheroidal.  The first encounters lead to
a bar-like phase that is rapidly heated into a more prolate
configuration.  As the stars and gas lose angular momentum to the dark
halo and perturbing galaxies, it becomes harder to physically remove
material from the galaxy.

This strong evolution of galaxies in clusters has seemed
counterintuitive to many of our colleagues prompting us to remark on
``harassment" in the broader context of collisional dynamics.  In
most stellar systems, the first sign of encounters is
``relaxation"---the redistribution of orbital velocities leading to
a Maxwellian distribution.  The relaxation time in a cluster of
galaxies is 100 times their lifetime.  However, the binding energy of
a small spiral galaxy is $\lsim 10^{-2}$ of its orbital energy and
$\lsim 10^{-3}$ of the orbital energy of an $L_*$
galaxy.  Internal energy transfers via galaxy-galaxy interactions provides an
important internal heating source with a timescale that is much
shorter than the relaxation time.  These circumstances are clearly
unique when we contrast them to other astrophysical environments where
encounters occur, {\it e.g.}  close stellar encounters in star
clusters (Press and Teukolsky 1977) or the interaction between a
proto-stellar disk and passing stars (Ostriker 1994).  In the former
case, the orbital kinetic energy is much less than the stellar binding
energy, whereas in the latter case the internal and orbital energies
are comparable.  These cases lead to an intuition about the timescales
of ``relaxation" versus internal heating in stellar systems that is
clearly false for small galaxies in clusters.

\section{The properties of harassed disk galaxies}

Table 1 displays a variety of data for the final states of the remnants.
Models 1 through 10 are the larger disks and
11 through 15 are the smaller ones.  Column 2 states
whether other galaxies were included
as perturbers within the cluster (the alternative being a 
completely smooth
cluster potential).  The
apocluster and pericluster radii of the initial disk's orbit are given
in column 3.  Columns 4---7 are the final masses of
the remnants: the fraction of stars that remain
bound to the final remnant $M_{*,f}/M_{*,i}$, the total mass of the
stars in the remnant in solar units $M_{*,f}/\msun$, the total mass of
dark matter within the tidal radius of the final remnant in solar
units $M_{d,f}/\msun$ and the ratio of total mass to the disk mass
inside the half-light radius of the remnant $M_{T,f}/M_{*,f}$.
Column 8 is the tidal radius of the remnant fit from the final
density profile of dark matter $r_t$ while column 9 is the effective
radius defined by the projected radius containing half of the light
$r_e$.  Column 10 is the mean column density of stars within $r_e$.
The three components of the velocity ellipsoid $\sigma_x$, $\sigma_y$
and $\sigma_z$ are shown in column 11.  Column 12 gives the rotation
velocity about the the three axes $nx$, $n_y$ and $v$.  The last two
columns define the shape of the three dimension figure, measured at
$2r_e$ using the ratio of the short to long axis $c/a$ and the
intermediate to long axis $b/a$.  If $c/a \sim b/a$, the object is
nearly prolate.  If $b/c \sim 1$, then it is oblate.

%*** with this definition of shapes, prolate = oblate
%*** what are nx n_y and v above?

\subsection{Evolution of the mass distribution}

The  dark halos were constructed out to the
tidal radius defined by the cluster potential 
at the galaxy's pericentric distance.
In the comparison runs without other perturbing galaxies, 
we find that some of the dark halo is stripped owing
to tidal shocks at pericenter, but the simulations with perturbers
demonstrate that  far more mass is
lost owing to encounters with perturbing galaxies.
At the start, model galaxies are dark matter dominated beyond about 2 disk
scalelengths.  After several billion years of evolution and
heating, they are frequently baryon-dominated at all radii. 
At a fixed radius in the galaxy, more dark matter can be lost than stars
because of the orbital distribution of the two components. 
The dark matter is on 
eccentric orbits that reach the tidal radius while disk 
material on circular orbits remains safely bound.
This evolution is shown in Figure 1(a) where we plot the
contribution to the effective rotation curve provided 
by the stars and dark matter separately. 
Figure 1(b) shows similar results but for one of the simulations with gas.

Figure 2 shows the initial and final mass distributions of the separate
components: stars,
gas, dark matter and the combined total.  We find that
25--75\% of the stars are stripped over about 5 Gyrs. When 
we include a gaseous disk, the stellar mass loss drops
significantly owing to the central density increase. We emphasize the
importance of the inner mass distribution in determining the evolution
of the system. If the central mass distribution can respond
adiabatically to the perturbation then the visible 
effects of harassment in our simulations are considerably weaker. 
However, we have not included a recipe for star formation so it is
unclear how sensitive the star-formation rate is to tidal
perturbations.

\subsection{Luminosity profiles} 

Figure 3(a) shows the time evolution of the stellar
density as the model galaxy moves on a circular orbit at 450
kpc without any perturbing galaxies. 
After 5 Gyrs of evolution, neither the density plot nor 
visual inspection shows much change.   In Figure 3(b), 
the same model galaxy is on an eccentric orbit with apocenter:pericenter
of 600:300 kpc, again
without any pertubing galaxies.  The first
pericentric passage causes a bar instability followed by an increase
in the inner stellar density.  The outer
stellar 
density profile doesn't show much change.  Figure 3(c) is the same
high resolution galaxy model as Figure 3(b) 
but includes the other cluster
galaxies as perturbers.  Now, the stellar density steepens 
significantly beyond $\sim$kpc as
mass is stripped from the outer disk.

If we follow a gaseous component then the stellar density profile 
evolves more strongly in the central region, but less strongly
beyond a few kpc. The gas collapses with the stars into a bar
configuration, but continues to cool and sink as it looses energy via
radiation and angular momentum loss to the stars and dark halo. The
contraction of the gas causes an adiabatic contraction of the stellar
distribution, leading to a power-law luminosity profile down to our
resolution limit.  The increase in binding energy due to the change in
central density makes it harder for perturbations to strip stars from
the galaxy. As a result, the outer luminosity profile evolves less
strongly than in the simulations without gas \--- see Figure 3(d).
We note here that the luminosity profiles are strongly affected
by the presence of a dissipative component.   The inclusion of star
formation and feedback would likely halt the collapse of the gas 
and resulting profile  would lie between the two extreme cases that
we examined.

Binggeli and Cameron (1991) provide luminosity profiles for several
hundred spheroidal and nucleated dwarfs (dE and dE,N respectively in
their notation). Most bright spheroidals show an inner extended excess 
of luminosity
over a King (1966) model ($\sim 1$ arcsecond $\equiv 100$ 
pc at the distance of the Virgo cluster). The 
latter class are distinguished by an 
additional ``stellar-like'' nulceus.
Most spheroidals fainter than $M_{_B}\sim-16$ show no
excess.

Figure 4 shows the surface brightness profiles of the harassed
remnants.  Our simulations were not designed to probe the inner 200
pc, but we do see a clear difference between our simulations that have
gas and those that do not. We ran models
that had identical mass distributions and orbits with and
without gas.  The global
evolution from a disk to a spheroid 
is comparable, but the inner density profiles change.  In
the ``no gas" case (dashed--dot line in Figure 4), the remnants are
spheroids without a large central luminosity excess.  
In the models ``with gas" (dashed
line in Figure 4), the torques of collisions together with the
cooling causes the gas to sink and form 
a density excess similar to the luminosity spike in the nucleated
bright
spheroidals (Binggeli and Cameron 1991).  Even without gas,
the profile often becomes steeper and the central density increases
compared to the initial disk.  This is not caused by two body
relaxation; the relaxation times in our simulation
are far too long.  (This was checked
by verifying that the 
stellar phase densities are not increasing.)  Some of our
harassed remnants resemble dS0 systems, the
lenticular class arising from the survival of a thick stellar
disk (Sandage \& Binggeli 1984).

\subsection{Flattening}

The shapes of harassed remnants become progressively rounder
as the heating continues.  After about 5 billion years of evolution
within the cluster the remnants without gas have
shapes between E3 to E5, e.g. somewhat flatter than the E3 shape that
is the mean for giant ellipticals.  The inclusion of gas tends to make the
final shapes less eccentric as the ensuing central concentration
causes a rounder potential in the inner parts. The increased central
density concentration might make it harder to support the box orbits
that are important in maintaining triaxial shapes (Lake and Norman
1983, Gerhard and Binney 1985).

The spheroidals in Virgo show the same trend as they are clearly
flatter than normal
ellipticals (Ryden and Terndrup 1994).  Binggeli and Popescu (1995)
separate the nucleated spheroidals and find that they are rounder than
the others having flattenings that are similar to giant ellipticals.

With our small exploration of parameter space we cannot study trends
in the harassed remnants.  However, there are several that one might
expect.  As the collisional heating progresses, 
the galaxies should become progressively larger and rounder creating a 
anti-correlation between surface brightness and flattening.

Figure 5 shows the evolution of the axial ratios in various 
simulations, owing 
to the ``tidal shocks'' at pericentric passage. 
(There is a slight additional heating of the disk owing
to the discreteness of the dark halo model.)  Figure 5(a) shows the same
galaxy on an eccentric orbit with no perturbers, but with the disk
orbiting within (dotted line) and perpendicular (solid line) 
to the orbital plane of the galaxy within the cluster.
The ratio of minor to major axis (c/a) measured at twice the effective
radii increases from
0.05 to $\sim$0.2 over 5 Gyrs and disks that
lie in the orbital plane are more strongly affected by the cluster
tidal field.

When the other perturbing galaxies 
are included, the disk heating is substantial and the
stellar shape evolves dramatically with time. Figure 5(b) shows
the evolution of c/a  in 4 separate simulations. The first
strong encounter drives a strong bar instability and the subsequent
evolution shows a steady increase in axial ratio towards 
``rounder'' shapes.  The final c/a values range from 0.25 to
0.75.
Again, the inclusion of a gaseous disk slows down the evolution
and the stellar configuration becomes less responsive to perturbations as
the central density increases (Figure 5(c)).

\subsection{Stellar kinematics and shapes} 

The stellar velocity dispersions of models on eccentric orbits without 
other perturbing galaxies 
increases slowly with time as tidal shocks from pericentric passage 
heat the disk.   
When perturbers are included
the velocity dispersions rise more quickly and tend to
level off or even start to decrease slowly with time as increasing mass loss
causes the galaxy to swell. These effects are shown in Figure 6 where we
plot the evolution of the stellar velocity dispersion.

The flattening of galaxies owes to either rotation or
anisotropic pressure.  For oblate systems that are flattened by
rotation, there is a simple relationship between the observed
flattening and $v/\sigma$ the ratio of the rotation velocity to the
line-of-sight dispersion (Binney 1978) shown as a solid line in 
Figure 7.
Lake (1983) introduced the parameter $\vstar$, the ratio of the
observed $v/\sigma$ to that required to produce the observed
flattening.  
In giant ellipticals, $\vstar \lsim 0.3$, while bulges
and compact dwarf ellipticals like M32 are flattened by rotation,
e.g. $\vstar \sim 1$ (Lake 1983, Davies \etal 1983).  There are very
few spheroidal galaxies with kinematical data, but the sample of six
compiled by Ferguson and Binggeli (1994) all have significant velocity
anisotropy with $\vstar \lsim 0.4$.  They are clearly not following the trend
shown for dwarf ellipticals, but their flattening may owe slightly more
to rotation than seen in the giant ellipticals.

The harassed remnants are triaxial ranging from nearly prolate to
nearly oblate \--- see Figure 7.  
In all cases, the shortest axis is the rotational
axis.  This leads to a variety of interesting kinematical tests
(Binney 1985).  When viewed down the intermediate axis, they show the
maximum amount of rotation and flattening ($c/a$ and
$(v/\sigma)_{max}$ in Table 1).  When viewed down the long axis, they
appear rounder with nearly the same value of $v/\sigma$ and the
maximum value of $\vstar$ ($b/a$, $(v/\sigma)_{max}$, and
$\vstar_{max}$ in Table 1).  Finally when viewed down the short or
rotation axis, they show a flattening equal to the ratio of the
intermediate axis to the long axis ($b/c$ in Table 1) and $\vstar \sim 0$.   
It is difficult to compare the complicated distribution derived by
sighting through arbitrary viewing angles with a sample of six observed
galaxies, but the observed trend of more rotation than giant ellipticals
but less than needed for rotational flattening is certainly true for
our remnants.

\subsection{The fundamental plane} 

In the multivariate space of galaxy properties (luminosity,
size, surface brightness, velocity dispersion, line strengths),
galaxies are observed to have highly correlated properties such that
they populate a ``fundamental plane" (Djorgovski and Davis 1987).
Extremely tight correlations exist for elliptical galaxies and bulges
that link them as a single family.  Spiral disks show their own
distinct correlations.  The data on spheroidals is sparse owing to the
difficulty of measuring the internal kinematics of such faint low
surface brightness galaxies.  There are some data between $-15 \gsim
M_B \gsim -17$ with a large scatter.  To define a plane, these data
are ``connected" to the observations of the dSphs ($-8 \gsim M_B \gsim
-11$) leaving a gap of nearly 5 magnitudes in luminosity.  A
departure from the giant E fundamental plane is clear, although 
the existence of a separate plane is not (Kormendy 1985, 
Ferguson and Binggeli 1994).

For normal ellipticals, the tight correlations over 2 decades in
luminosity are remarkable given the many and varied ways that galaxies
are thought to form and evolve.  Our proposal that harassment creates
spheroidals leads to qualitative and quantitative statements regarding
the region of parameter space they inhabit in projections designed to
display a fundamental plane.  Since their progenitors are disks, we
expect more similarity to disk properties than to ellipticals.  
As they evolve from disks to spheroids, 
the central properties typically change by a factor of 2,  rather than the
orders of magnitude that might be required to reach the plane defined
by elliptical galaxies.
Since the remnants form by stochastic reshaping and stripping,
it would be remarkable if
they populated a slender plane.   However, survival of the final 
remant imposes some strong contraints, as one might easily guess
from the fragility of the initial disks.  
From our limited number of simulations
we find that the final stellar configurations have half light radii up
to 50\% smaller than the progenitor and a central surface brightness that
is 2---3 times larger. The scatter in surface brightness at a fixed $r_e$ 
is as large as a factor of 2.

\subsection{Stellar populations}

Ellipticals have old stellar populations that are 
well-modeled by a single burst of star-formation (c. f. Bender,
Burstein and Faber 1993).  Spheroidals have
intermediate age populations betraying several prominent
episodes of star-formation (Held and Mould 1994).  Since ellipticals and
spheroidals are both deficient in gas, it's something of a surprise
that the least massive systems retained their gas for a late epoch of
star-formation.  However, this follows naturally from our linkeage
of the spheroidals to the galaxies forming stars in
clusters at $z \sim 0.4$ where the spectral data imply
multiple bursts of star formation (Couch \etal 1994, 
Barger \etal 1996).  We consider radial gradients in \S 3.9.

\subsection{Luminosity function of spheroids produced by harassment} 

The bright end turnover of the spheroidal luminosity function occurs
at $\gsim$ 4 magnitudes fainter than
$M_*$, there are exceedingly few that are brighter
than 3 magnitudes below $M_*$.  Binggeli \etal (1988) note
that the faint end of the luminosity function is dominated entirely
by spheroidals in clusters, while Sd and Im galaxies are dominant in
lower-density environments.  They state that differences in the slope
at the faint end might be explained entirely by the variations in the
ratio of spheroidals to Sd-Im galaxies.  Details of the faint end
luminosity function are uncertain owing to the myriad of selection
effects that plague the inventory of faint, low surface brightness
objects (Ferguson and McGaugh 1994), so we will focus our attention on
the break at the bright end.

In our work on the collision and tidal heating of spherical systems
(Moore, Katz and Lake 1996), we showed that {\sl harassment has
virtually no effect on a system as dense as a giant elliptical galaxy
or a spiral bulge.}  As a result, only purely disk galaxies can be
turned into spheroidals.  Hence, the maximum luminosity of a purely
disk system that falls into the cluster will determine the maximum
brightness of the resulting dSph. The luminosity function of 
galaxies is strongly type
specific and pure disk galaxies (Sc or later) with circular
velocities greater than $\sim 160 \kms$ are exponentially rare (Tully,
Mould and Aaronson 1982, Meisels 1983, Binggeli \etal 1988).  The
cutoff for Sd-Im galaxies occurs at roughly $125 \kms$.  In terms of
magnitudes fainter than $M_*$, these cutoffs are roughly 1.6 and 2.8 magnitudes
respectively.

The fate of our model disks with a circular velocities of $160 \kms$
should be representative of the evolution of the brightest purely disk
galaxies that enter the cluster.  They lose between 25\% and 75\% of
there stars ({\it i.e.} up to a magnitude) and then presumably fade at least a
magnitude as the stellar population ages.  Taken together, the break
at the bright end of the Sc or Sd-Im luminosity functions should create
a comparable break in the spheroidal luminosity function that lies 
$\sim$4 magnitudes below \l*.

\subsection{Mass-to Light Ratios, Connection with the dSph Milky Way 
Satellites } 

As we mentioned in \S 3.4, kinematical data on the spheroids is sparse
with large scatter in the properties of the handful of galaxies
observed between $-15 \gsim M_B \gsim -17$.  The mass to light ratios
of the small sample of cluster spheroids ranges from 2 to 8 (Petersen
and Caldwell 1993, adjusted for our adopted Hubble constant).  Column
7 of Table 1 shows the ratio of total mass to stellar mass
$M_{T,f}/M_{*,f} \sim 1.5-2$ within the effective radius $r_e$.  To
get a mass-to-light ratio for the remnant, one multiplies the expected
$M/L$ for the stellar population by this quantity.  For these
relatively low metallicity systems, one might expect $M/L_* \sim 3 \pm
1$, leading to global values of $3 < M/L < 8$ for our remnants, in
good agreement with the observations.

The only hope of discerning trends in quantities like $M/L$ is to
connect the scattered cluster points between $-15 \gsim M_B \gsim -17$
to the even more broadly scattered points defined by local dSphs at
magnitudes ($-8 \gsim M_B \gsim -11$).  The main conclusion drawn is
that the mass-to-light ratio is increasing inversely as a power of
luminosity, typically $M/L \propto L^{-0.4}$ owing to the $M/L \gsim
30$ of Draco and Ursa Minor (Armandroff, Olszewski and Pryor 1995).

What are we to make of the high $M/L$ values of the dSph galaxies?
Could Draco and Ursa Minor be remnants of harassment?  Earlier, we
stressed that harassment is a collective process, phase space
densities of dissipationless material do not increase.  Therefore a
progenitor Irr galaxy harassed to make Draco or Usra Minor must have
high central densities and phase densities of dark matter.  
Such progenitors may well have existed (Lake 1990a).  
The extreme dwarf irregular GR8 (Carignan, Beaulieu
and Freeman 1990) has a central dark matter density of $\sim 0.1 \msun
{\rm pc}^{-3}$ which is only a factor of two less than what is required in
Draco and Ursa Minor (Lake 1990b).  GR8 is the only such Irr that has been
measured while Draco and Ursa Minor are the most extreme dSph
galaxies.  Further, GR8 has a galactocentric radius of 1 Mpc, 15 times
that of the other two.  Despite this gross difference in location
and its implications for survival, GR8 is close enough to Draco and
Ursa Minor 
that it's not unreasonable to postulate appropriate Irr progenitors.

If harassment turned dwarf irregulars like GR8 into dSphs like Draco
and Ursa Minor, we can scale our cluster simulations and timescales to
determine the nature of the harassing perturbers in the early Milky
Way.  The virial mass and radius of an object scale as $R_{virial}
\propto v_{circ}$ and $M_{virial} \propto v^3_{circ}$.  Given the
Milky Way's circular velocity of $220 \kms$, we infer $M_{MW, virial}
= 2 \times 10^{-3} M_{Coma, virial}$ and $R_{MW, virial} = 0.15
R_{Coma, virial} = 220$ kpc.  The dSph galaxies have masses of $\gsim
10^7$ (Lake 1990b) which is $\sim 10^{-3}$ of the mass of the remnants
in Table 1.  The galactocentric radii of the Draco and Ursa Minor are
70 kpc or $0.32 R_{MW, virial}$ while the galaxies we simulated had
guiding centers of $0.3 R_{Coma, virial}$.  A galaxy similar to GR8
would be harassed into something resembling Draco and Ursa Minor in 3
Gyrs if the Milky Way had $N \sim 30$ perturbing lumps with mass
$M_{MW,pert} \sim 4 \times 10^8 \msun$.  More generally, the timescale
for harassment of GR8 into a dSph is roughly $\tau \sim 3 Gyr (N/30)^{-1}
(M_{MW,pert}/(4 \times 10^{8}))^{-2}$.  Models of the early Milky Way
often appeal to substructures of comparable mass and number (Searle
and Zinn 1978).

We note that Ursa Minor and Draco are well within the virial radius
of the Milky Way while the Dwarf Irregular GR8 is clearly well
beyond it.  Harrassment would not be able to explain dSphs that
were members of the Local Group but
clearly not satellites of the Milky Way or M31.  At a distance
of 230 kpc, Leo I would appear to be the greatest challenge
(Zaritsky \etal 1989).  However, it's observed to move outward
with a radial velocity of $178 \kms$,  so it could have been
closer to us than Ursa Minor and Draco just a billion years ago.

\subsection{Correlations with clustercentric radius/density}

There are two effects that can cause correlations with clustercentric
radius---the dynamics of harassment change
with radius and hierarchal clustering can produce inherent
differences between the center and outer reaches of the cluster.

Harassment depends on collisional frequency, the strength of 
individual collisions and the strength of the cluster's tidal
field.
If the dark halos of galaxies scale with radius as the tidal limit,
then the heating rate owing to collisions 
is independent of cluster radius (the collision rate is proportional
to galactic density which scales as $R_{clus}^{-2}$, but this is balanced
by the collisional strength since the tidally limited 
mass of individual galaxies 
varies linearly as $R_{clus}$ and the strength of the collision
scales with the square of the perturber's 
mass).  In the impulse approximation, the heating rate is even constant
for any galaxy that finds itself inside a larger 
virialized structure---independent of the size of the structure.
However, the impulse approximation will clearly break down at large
radii since collisions in a rich cluster cease to be impulsive if
the impact parameter is greater than $\sim 100$ kpc.   (The impulse approximation
breaks down faster if the galaxy is in a smaller virialized structure
with lower relative velocities).  

The cluster's tidal field  becomes exceedingly important at the center.
Modeling both a small galaxy and a cluster as an isothermal sphere, we
find that the tidal radius $R_{tidal}$ of a galaxy with a
dispersion $\sigma_{gal}$ with pericentric radius $r_{peri}$
in a cluster of dispersion $\sigma_{clus}$ is given by:

$R_{tidal}= 5 {\rm kpc} \bigl({{r_{peri}} \over {100 kpc}}\bigr) 
\bigl({{\sigma_{gal}} \over {50 \kms}}\bigr)
\bigl( { {\sigma_{clus}} \over {1,000 \kms} } \bigr)^{-1}$

\noindent Recall that a galaxy seen at a distance of $r_{now}$ in a cluster has
a typical pericenter of $r_{now}/3$, so the typical object with mean
orbital radius 
of $r_{now}$ will have a mean tidal radius:

$R_{tidal,mean} = 1.67 {\rm kpc} 
\bigl({{r_{now}} \over {100 kpc}}\bigr) \bigl({{\sigma_{gal}} \over {50 \kms}}\bigr)
\bigl( { {\sigma_{clus}} \over {1,000 \kms} } \bigr)^{-1}$

\noindent As a result, the smaller spheroidals will be destroyed in 
the innermost part of the cluster.  

Radial gradients of the spheroidal populations may provide some interesting
tests of our model.  However, one important effect will be common to all
models---independent of
how the spheroidals might have formed, global tides
will coerce the the lowest density 
(or surface brightness) objects into the diffuse stellar background.
We expect that most observed radial correlations will be projections
of the fundamental correlation between density and survivability.
For example,
Secker (1996) looked for a radial gradient in spheroidal colors 
in the Coma cluster.  He fit a power law to his observed distribution
$(B-R) \propto R_{proj}^{0.08 \pm 0.02}$, but the data is flat except for
a point at 2 arcminutes (40 h$^{-1}$kpc) that is redder by $\Delta (B-R)
\sim 0.1$ than the eight outer points.  The detailed scatter of the
points in plots makes it appear that the inner bin is deficient in bluer
points rather than a shifting of the entire distribution with radius. 
Secker remarks that there is a strong color-luminosity relationship, but 
quotes theoretical reasons against luminosity segregation that he felt were
so strong that he didn't examine his own data.
By contrast, Bernstein \etal (1995) clearly see a
deficiency of low luminosity galaxies in the inner few arc
minutes of this cluster.      

Finally, radial gradients in properties can arise from the history
of the cluster's hierarchical formation via
mergers and accretion.   In general, 
the material at the cluster center is older
than the material in the outer regions.  
Hence, the objects in the cluster center  experience 
harassment at earlier times and for longer durations. 

\noindent From these general principles, we expect that:
\begin{itemize}
\item{} no pure disk galaxy can survive transformation within about half of the
virial radius of a large cluster
\item{} spheroidals with small pericentric radii 
would be destroyed by tides alone
\item{} the look-back time to when spirals were turned into spheroidals
and the diffuse light was liberated from galaxies increases toward the
center of the cluster. 
\end{itemize}
\noindent This leads to a variety of observable effects:
\begin{itemize}
\item{} only the densest spheroidals survive in the inner parts of the galaxies,
creating a paucity of the low luminosity spheroids in the central region
as seen by Bernstein \etal (1995)
\item{} since survival depends on density, any correlations
between density and color/metallicity  (including indirect correlations
owing to luminosity or binding energy) will create color gradients in
the surviving ensemble of spheroidals as seen by Secker (1996)
\item{} there will be an age gradient in the diffuse light with
the central regions being older than the inner regions, but
color gradients will depend on metallicity and the
central diffuse light will also include the ``splashes" of
cannibalism making specific predictions difficult (the previous 
``bullet" captures the dominant effect)
\item{} nucleated spheroidals will always be more robust than those without
a dense nucleus, therefore
the fraction of nucleated spheroids will increase towards the center 
({\it c.f.} van den Bergh (1986) and Binggeli \etal (1987))
\item{} the heating is greatest in the center leading to lower surface
brightnesses and rounder spheroidal galaxies, but the detailed distribution
with radius also depends on the survival of such galaxies 
\item{} the selective destruction of non-nucleated
spheroidals with small pericentric
radii will lead to a central velocity ellipsoid that is 
deficient in radial orbits causing a dip in the cluster's 
line-of-site velocity 
dispersion as measured by the non-nucleated spheroidals
\item{} nucleated spheroidals will show a higher central velocity dispersion
than non-nucleated
\item{} spiral disks seen in the central regions of clusters will 
be caused by projection leading to velocity fields that are not
virialised as in the case of the Virgo cluster's ``backside infall"
(deVaucouleurs 1961, Tonry, Ajhar and Luppino 1990)
or exhibiting nearly free expansion as seen in the Coma
cluster (Bernstein \etal 1994) 
\item{} in the outer parts of the clusters, the spirals on
radial orbits are transformed faster than those on nearly
circular orbits as seen in the analysis of gas rich spirals
by Dressler (1986)

\end{itemize}

\subsection{Clustering properties} 

In the Introduction, we noted that several prior models ran
afoul of clustering properties.  Is harassment any better?
It can clearly explain some of the broad trends.
Vader and
Sandage (1991) stressed that the spheroidals are the most strongly
clustered of all galaxies.  
Their frequency
increases sharply with density of the environment.  The Virgo cluster
has approximately 5 times as many dwarfs per giant as poor groups and
this ratio correlates with the group/cluster velocity dispersion as
expected in our model (Ferguson and Sandage 1991).
Finally, Binggeli, Tarenghi and Sandage (1990) find that  ``field
spheroidals"  are companions to 
giant field ellipticals.  They conclude that their phenomenology would
be explained by a mechanism that triggered the transformation
of Irr to spheroidals in high density environments, which is exactly
what harassment does.
We stress that harassment 
occurs in any virialized environment with significant
``lumps" to act as perturbers.  This is certainly the case for
clusters of galaxies and for all current models
of elliptical galaxy formation  (\S I).

%**** Do they only become rounder or do they become more oblate along
%the way, do the rounder ones get closer to being rotationally
%flattened?  Does flattening correlate with central surface brightness?

\section{Conclusion}

We have followed the evolution of small disk galaxies in clusters 
using realistic simulations of the combined action of rapid encounters 
with cluster galaxies and global tidal heating---``galaxy harassment".
We find that the first 
shocks create the population of distorted spiral galaxies
seen in HST images $z \sim 0.4$ cluster.  Over a period of a few Gyr,
these galaxies evolve into remnants that have surface
density profiles, shapes, kinematics, stellar populations and
clustering properties that qualitatively 
match the spheroidal systems in present-day
clusters.  The assumptions and parameters of our theory are minimal.
It relies only on gravitational interactions with the global cluster
potential and other galaxies.  The strength of these effects is well
determined by the observed properties of the clusters and their
brighter members.  Non gravitational effects such as ram pressure
stripping, wind driven expulsion of gas or other hydrodynamic
interactions can speed the timescale for the evolution of the gas, but
will not sufficiently alter the stellar disk of the galaxies observed
at $z \sim 0.4$.  Further, the harassment timescale matches the
interval of multiple starbursts that is inferred in clusters at $z
\sim 0.4$ (Couch \etal 1994).  This provides a limit on the magnitude
of additional hydrodynamic effects, assuring us that harassment must
be the dominant mechanism.

In this paper, we have focused on the remnants.  However, it is clear
that roughly 50\% of the stars escape into the intracluster medium.
They do so within debris tidal streams and thin tails that gradually
disperse.  A study of these low surface brightness features and the
general diffuse light in clusters will be the subject of a future
paper.

\acknowledgments

We would like to thank Thomas Quinn for numorous discussions and his
help producing {\sl Galaxy Harassment---The Movie}.
This research was supported by NASA's High Performance Computing and
Communications Earth and Space Sciences Program,  Astrophysical
Theory Program and Long Term Space Astrophysics Program.
Ben Moore is a Royal Society Research Fellow.

\baselineskip=8pt

\clearpage
\vskip 1.0truein

\noindent{\bf References}

%\doublespace

\pp Armandroff, T. E., Olszewski, E. W. and Pryor, C. 1995, {\it
A.J.}, {\bf 110} , 2131.

\pp Bardeen, J.M., Bond, J.R., Kaiser, N. and Szalay, A.S.  {\it
Ap.J.}, {\bf 304} , 15.

\pp Barger, A.J., Aragon-Salamanca, A., Ellis, R.S., Couch, W.J., Smail, I.
\& Sharples, R.M. 1996, {\it M.N.R.A.S.}, {\it 279}, 1.

\pp Barnes, J. 1988, {\it Ap.J.}, {\bf 331}, 699.

\pp Barnes, J. and Efstathiou, G. 1987, {\it Ap.J.}, {\bf 319}, 575.

\pp Bender, R., Burstein, D. and Faber, S. M. 1993, {\it Ap. J.}, {\bf
411}, 153.

%\pp Bender, R. and Nieto, J.-L. 1990, {\it Astr. Ap.}, {\bf 239}, 97.

\pp Bernstein, G.M., Guhathakurta, P..  Raychaudhury, S.. 
Giovanelli, R., Haynes, M. P.,  Herter, T. and Vogt, N. P. 1992
{\it A.J.}, {\bf 107}, 2307.

\pp Bernstein, G.M., Nichol R.C., Tyson J.A., Ulmer M.P. \& Wittman D.
1995, {\it A.J.}, {\bf 110}, 1507.

\pp Binggeli, B. and Popescu, C.C. 1995, {\it Astr. Ap.}, {\bf 298},
63.

\pp Binggeli, B. and Cameron, L.M. 1993, {\it Astr. Ap. Suppl..}, {\bf
98}, 297.

\pp Binggeli, B. and Cameron, L.M. 1991, {\it Astr. Ap.}, {\bf 252},
27.

\pp Binggeli, B., Sandage, A. and Tammann G.A.  1988, {\it
Ann. Rev. Astr. Ap.}, {\bf 26}, 509.

\pp Binggeli B., Tammann, G.A. and Sandage, A. 1987, {\it A.J.}, {\bf
94}, 251.

\pp Binggeli, B., Tarenghi, M. and Sandage, A. 1990, {\it A.J.}, {\bf
228}, 42.

\pp Binney, J. 1978, {\it M.N.R.A.S.}, {\bf 183}, 501.

\pp Binney, J. 1985, {\it M.N.R.A.S.}, {\bf 212}, 767.

\pp Butcher, H. and Oemler, A. 1978, {\it Ap.J.}, {\bf 219}, 18.

\pp Butcher, H. and Oemler, A. 1984, {\it Ap.J.}, {\bf 285}, 426.

\pp Byrd, G. and Valtonen, M. 1990, {\it Ap.J.}, {\bf 350}, 89.

\pp Carlberg, R. G. 1986, {\it Ap.J.}, {\bf 310}, 593.
 
\pp Carignan, C., Beaulieu, S. and Freeman, K. C. 1990, {\it Ap.J.},
{\bf 332}, L33.

\pp Couch, W.J., Ellis, R.S., Sharples, R. and Smail, I. 1994, {\it
Ap.J.}, {\bf 430}, 121.

%\pp Cowie, L. L. and Songaila, A. 1977, {\it Nature}, {\bf 266}, 501.

\pp Davies, R. L, Efstathiou, G., Fall, S. M., Illingworth, G. D. and
Schechter, P. 1983, {\it Ap.J.}, {\bf 292}, 371.

\pp Dekel, A. and Silk, J. 1986, {\it Ap.J.}, {\bf 303}, 39.

\pp Djorgovski, G. and Davis, M. 1987, {\it Ap.J.}, {\bf 313}, 59.

%\pp Dressler, A 1980, {\it Ap. J.}, {\bf 236}, 351.

\pp Dressler, A 1986, {\it Ap. J.}, {\bf 301}, 35.

\pp Dressler, A. and Gunn, J.E. 1983, {\it Ap.J.}, {\bf 270}, 7.

\pp Dressler, A, Oemler, A., Butcher, H. and Gunn, J.E.  1994a, {\it
Ap.J.}, {\bf 430}, 107.

\pp Dressler, A., Oemler, A., Sparks, W.B. and Lucas, R.A. 1994b, {\it
Ap.J.Lett.}, {\bf 435}, L23.

\pp Evrard, A.E. 1991, {\it M.N.R.A.S.}, {\bf 248}, 8p.

\pp Faber, S.M. and Jackson, R. 1976, {\it Ap. J.}, {\bf 204}, 668.

\pp Ferguson, H.C.  and McGaugh, S. S. 1994 {\it Ap. J.}, {\bf 440},
470.

\pp Ferguson, H.C. and Sandage, A. 1991, {\it A.J.}, {\bf 101}, 765.

\pp Ferguson, H.C. and Binggeli, B. 1995, {\it Astr. Ap. Rev.}, {\bf
vol. 6}, 67.

\pp Fisher, J. R. and Tully, R. B. 1981, {\it Ap. J. Suppl.}, {\bf
47}, 139.

\pp Gerhard, O. E. and Binney, J. J. 1985, {\it M.N.R.A.S.}, {\bf
216}, 467.

%\pp Ghigna, S., Moore, B., Governato, F., Lake, G., Quinn, T. \& Stadel, J.
%1998, {\it Ap.J.}, submitted.
 
\pp Gott, J. R. III 1977, {\it Ann. Rev. Astr. Ap.}, {\bf 15}, 235.

\pp Griffiths, R.E. \etal 1994, {\it Ap.J.Lett.}, {\bf 435}, L19.

\pp Gunn, J.E. 1982, in {\sl Astrophysical Cosmology, Proceedings of
the Vatican Study Week on Cosmology and Fundamental Physics},
ed. M.S. Longair, G.V. Coyne and H.A. Bruck, (Vatican City: Pontifcia
Academia Scientiarum).

\pp Held, E.V. and Mould, J.R. 1994, {\it A.J.}, {\bf 107}, 1307-19.

\pp Hernquist, L. 1993, {\it Ap.J.Suppl.}, {\bf 86}, 389.

\pp Hoyle, F. 1945, {\it M.N.R.A.S.}, 105, 288.

\pp Hubble, E.  1936, {\sl The Realm of the Nebulae}, (Oxford: Oxford
University).

\pp Icke, V. 1985, {\it Astr. Ap.}, {\bf 144}, 115-23.

\pp Jeans, J. 1938, {\sl Cosmology and Cosmogony}, (Oxford: Oxford
University).
 
\pp Katz, N. 1992, {\it Ap. J.}, {\bf 391}, 502.

\pp Kauffmann, G. 1995, {\it M.N.R.A.S.}, {\bf 274}, 153.

\pp King, I. R. 1966, {\it A. J.}, {\bf 71}, 64.

\pp Kormendy, J. 1985, {\it Ap.J.}, {\bf 295}, 73.

\pp Lake, G. 1983, {\it Ap. J.}, {\bf 264}, 408.

\pp Lake, G. 1989, {\it A. J.}, {\bf 97}, 1312.
 
\pp Lake, G. 1990a, {\it Ap. J. Letters}, {\bf 356}, L43.

%\pp Lake, G. 1990b, {\it Ap. J. Letters}, {\bf 364}, L1.

\pp Lake, G. 1990b, {\it M.N.R.A.S.}, {\bf 244}, 701.

\pp Lake, G. and Norman, C.  1983, {\it Ap. J.}, {\bf 270}, 51.

\pp Lake, G. and Dressler, A. 1986, {\it Ap. J.}, {\bf 310}, 605.

\pp Lake, G. and Carlberg, R. G. 1986a, {\it A.J.}, {\bf96}, 1581.

\pp Lake, G. and Carlberg, R. G. 1986b, {\it A.J.}, {\bf96}, 1587.

\pp Lake, G., Katz, N. and Moore, B. 1996, {\it Ap.J.}, this volume.

\pp Larson, R. B. 1974, {\it M.N.R.A.S.}, {\bf 169}, 229.

\pp Meisels, A. 1983, {\it Astr. Ap.}, {\bf vol. 118}, 21.

%\pp Merritt, D. 1983, {\it Ap.J.}, {\bf 264}, 24.

\pp Merritt, D. 1985, {\it Ap.J.}, {\bf 289}, 18.

\pp Miller, R.H. 1988, {\it Comment. Astrophys.}, {\bf 13}, 1.

\pp Moore, B., Katz, N. and Lake, G. 1996, {\it Ap.J.}, {\bf 457},
455 (MKL).

\pp Moore, B., Katz N., Lake G., Dressler, A. and Oemler, A. 1996, {\it
Nature}, {\bf 379}, 613.

%\pp Nulsen, P. E. J. 1982, {\it  M.N.R.A.S.}, {\bf 198}, 1007.

\pp Oemler, A., Dressler, A. and Butcher, H. R. 1996, {\it Ap.J.},
submitted.

\pp Ostriker, E.C. 1994, {\it Ap.J.}, {\bf 424}, 292.

\pp Ostriker, J.P. 1980, {\it Comm. Astr. Ap.}, {\bf 8}, 177.

\pp Ostriker, J.P. and Hausman M.A. 1977, {\it Ap.J.Lett.}, {\bf 217},
L125-9.

\pp Peterson, R. C. and Caldwell, N. 1993, {\it A. J.}, {\bf 105},
1411.

\pp Press, W.H. and Teukolsky, S.A. 1977, {\it Ap.J.}, {\bf 213}, 183.

\pp Richstone, D.O. and Malmuth, E.M. 1983, {\it Ap.J.}, {\bf 268},
30.

\pp Ryden, B.S. 1988, {\it Ap.J.}, {\bf 329}, 589.

\pp Ryden, B.S. and Terndrup, D.M. 1994, {\it Ap.J.}, {\bf 425}, 43.

\pp Sandage, A. 1961, {\sl The Hubble Atlas of Galaxies}, (Washington
D.C.: Carnegie Institution of Washington).

\pp Sandage, A., Binggeli, B. \& Tammann, G.A. 1985, {\it A.J.}, {\bf 90}, 1759.

\pp Sandage, A. \& Binggeli, B. 1984, {\it A.J.}, {\bf 89}, 919.

\pp Secker, J. 1996, {\it Ap.J.Lett.}, {\bf 469}, L81.

\pp Searle, L. and Zinn, R. 1978, {\it Ap.J.}, {\bf 225}, 357.

\pp Thompson, L.A. and Gregory, S.A. 1993, {\it A.J.}, {\bf 106},
2197.

\pp Tonry, J. L.,  Ajhar, E. A.  and Luppino, G. A. 1990, 
{\it A.J.}, {\bf 100}, 1416.

\pp Toth, G. and Ostriker, J. P. 1992, {\it Ap.J.}, {\bf 389}, 5.

\pp Toomre, A. 1964, {\it Ap.J.}, {\bf 139}, 1217.

\pp Toomre, A. 1977, In {\sl The Evolution of Galaxies and Stellar
Populations}, ed. B.M. Tinsley and R.B. Larson, p. 401, (New Haven:
Yale University Observatory).

\pp Tully, R. B., Mould, J. R. and Aaronson, M. 1982, {\it Ap.J.},
{\bf 257}, 527.

\pp Vader, P. 1991, {\it Ap.J.}, {\bf 305}, 669.

\pp Vader, P. and Sandage, A. 1991, {\it Ap.J. Letters}, {\bf 379},
L1.

\pp Valluri, M. 1993, {\it Ap.J.}, {\bf 408}, 57.

\pp Valluri, M. and Jog, C. J. 1991, {\it Ap.J.}, {\bf 374}, 103.

\pp Vaucouleurs, G. de. 1961, {\it Ap.J.Suppl.}, {\bf 6}, 213.

\pp van den Bergh, S. 1982, {\it Pub. A. S. P}, {\bf 94,459}.

\pp van den Bergh, S. 1986, {\it A.J.}, {\bf 91}, 271

\pp van der Marel, R. P. 1991, {\it M.N.R.A.S.}, {\bf 253}, 710-26.

\pp Zaritsky, D.,  Olszewski, E. W.,  Schommer, R. A.,  
Peterson, R. C. and Aaronson, M. 1989, {\it Ap.J.}, {\bf 345}, 759.

\pp Zepf, S. E. and Ashman, K. M., {\it M.N.R.A.S.}, {\bf 264}, 611.

\baselineskip=14pt

\clearpage
\centerline{\bf Figure Captions}
\bigskip

%\doublespace

\noindent{\bf Figure 1.} \ \ \ (a) The dashed curves are the effective
circular velocity at a radius $r$ from 
the mass distribution of the dark matter alone.  Likewise the solid
curves show the contribution to the effective rotation curve from
the stars.  These are both plotted for the initial and final states.  This
model is does not include SPH.  (b) As (a) but shows the effect of
including a gaseous disk and more time snapshots are plotted.

\bigskip

\noindent{\bf Figure 2.} \ \ \ The effective circular velocities 
at a radius $r$ from the center of the model
galaxy.   Each panel shows the contributions to $V_c$ from the separate
components, stars, gas, dark matter and total mass. This model is the
same that is plotted in 3(b). The solid  and dotted curves show the 
rotation curves at the initial and final
times respectively.

\bigskip

\noindent{\bf Figure 3.} \ \ \ The time evolution of the 
luminosity density as the model galaxies
evolve in the ``Coma" cluster:  (a) A circular orbit at 450 kpc
with no perturbers,  (b) An eccentric orbit with apocenter:pericenter
of 600:300 kpc and no
perturbers,  (c) As (b) but with perturbers included,  (d) As (c) and
including a full SPH treatment of a gaseous disk in the harassed
galaxy.

\bigskip

\noindent{\bf Figure 4.} \ \ \ The initial surface brightness 
profile of the Sd galaxy is shown as a solid line. The dashed and
dot--dash lines show the projected luminosity profiles of models with
and without a gaseous disk after 3 Gyrs of evolution in the cluster
with perturbing galaxies.

\bigskip

\noindent{\bf Figure 5.} \ \ \ The time evolution of the axial ratios
of the stellar component of various models:  $c/a$ is the axis ratio
between the long and short axes and $b/a$ is the ratio between the
intermediate and short axes, in each case measured at twice the
effective radii, $r_e$. Each panel shows the evolution of several
simulations: (a) eccentric orbit with no perturbers. The solid line is
a model in which the plane of the disk is not in the orbital plane and
the dotted line is a model that orbits in the plane defined by its
disk.  (b) eccentric orbits of four separate simulations that include
perturbers and random disk orientations, (c) eccentric orbit with
perturbers and including SPH treatment of the galaxy disk.

\bigskip

\noindent{\bf Figure 6.} \ \ \ Time evolution of the 1-d
velocity dispersion of the stars projected along each of the major, minor
and intermediate axis of the model: (a) the model
galaxy is on an eccentric orbit with no perturbers and the disk rotates
out of the orbital plane of the galaxy, (b) as (a) but the disk rotates
in a plane defined by the orbit of the galaxy in the cluster, (c) as (a)
but including perturbers, (d) as (a) but including perturbers and a gaseous
disk. The dashed lines show results from a separate simulation.

\bigskip

\noindent{\bf Figure 7.} \ \ \  The final shapes of the 11 
harassment simulations (perturbers included) plotted against
$v/\sigma$ ($v$ is the rotation velocity and
$\sigma$ is the projected velocity dispersion).  Each model is shown by
number for 3 projections: viewed down the 
long axis (squares), intermediate axis (circles) and short axis (numbers
only).   The projected flattening is measured at $2r_e$.
The line shows the theoretical curve for rotationally flattened
oblate spheroids. 

\bigskip

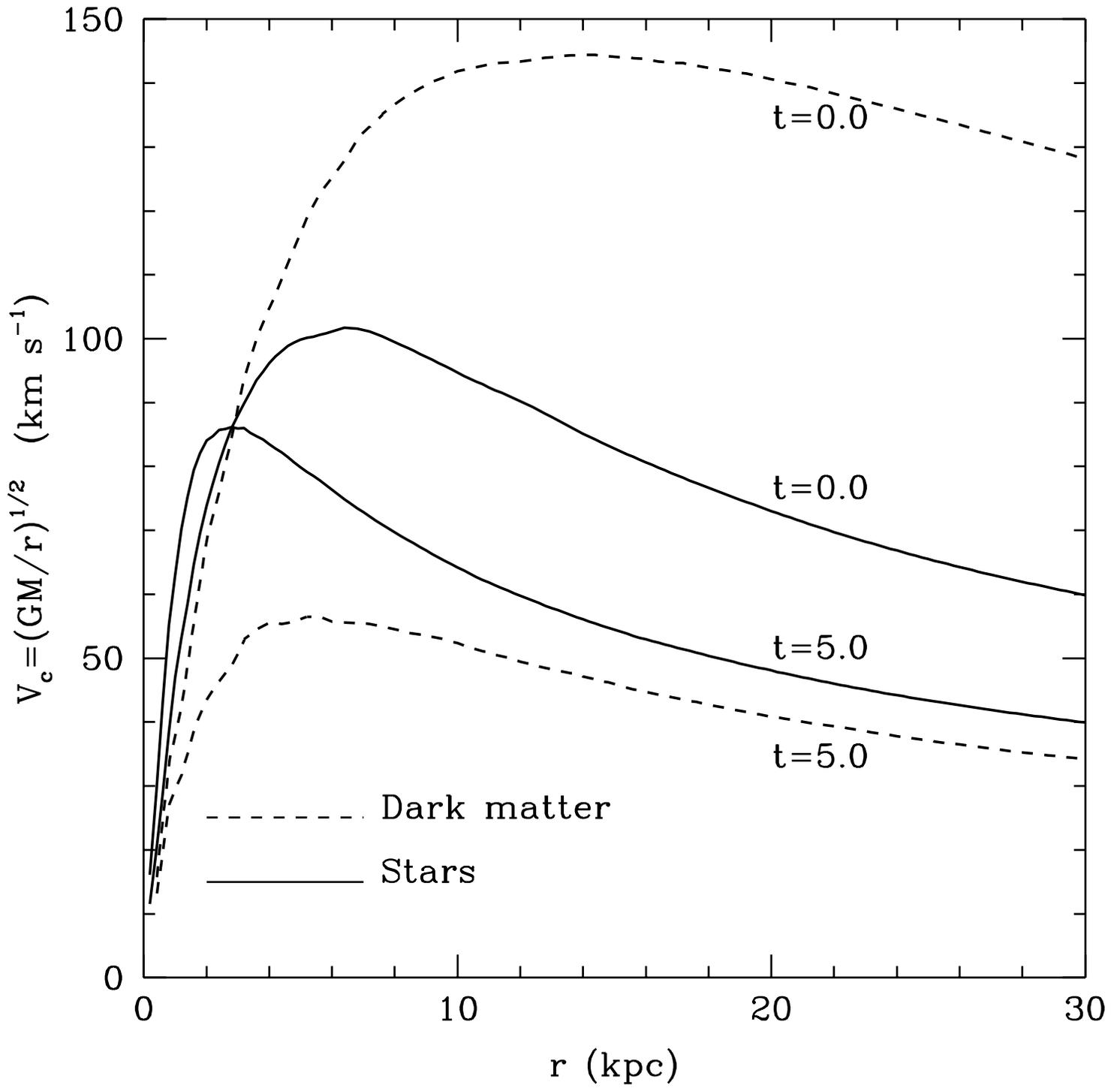
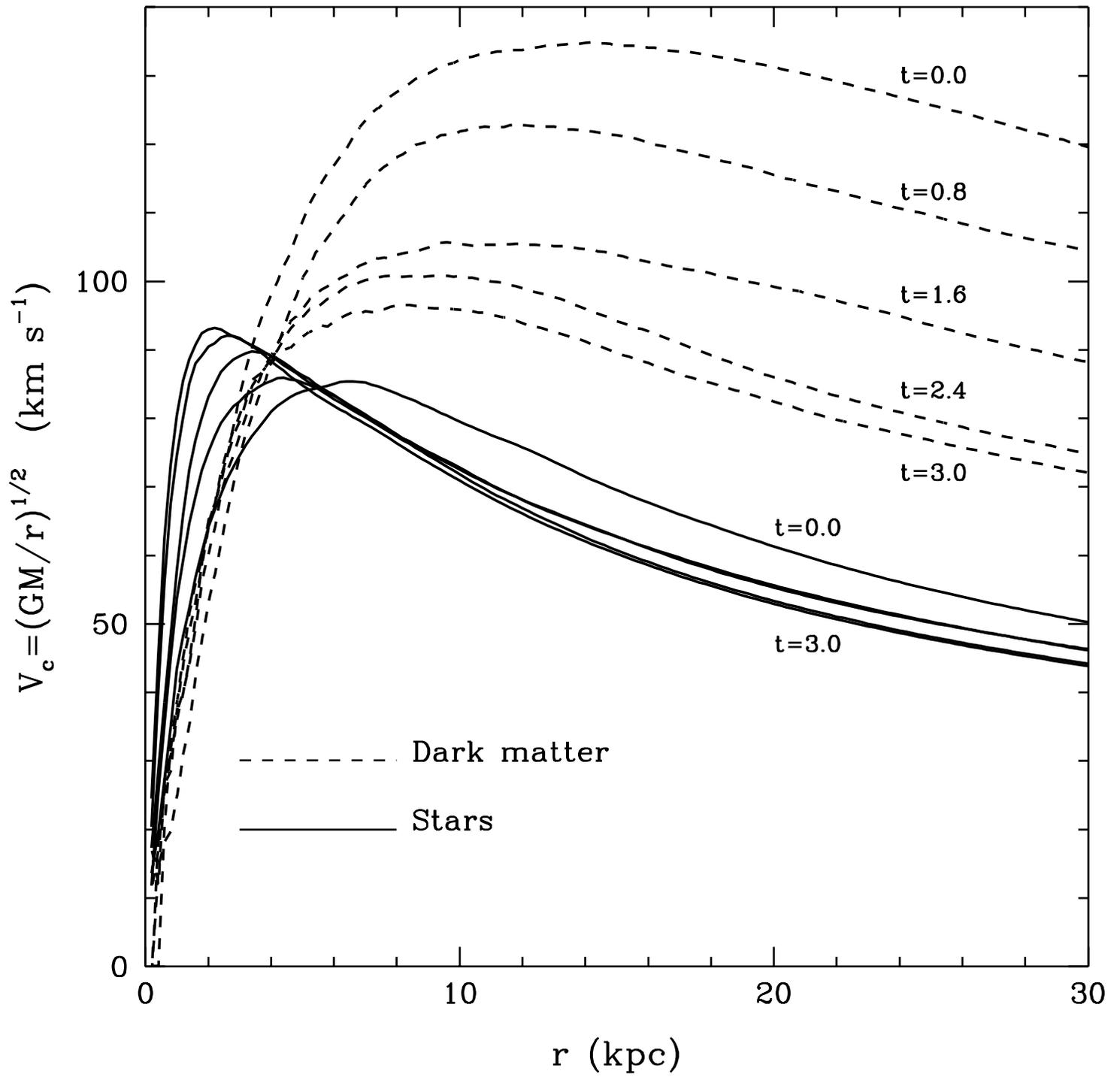
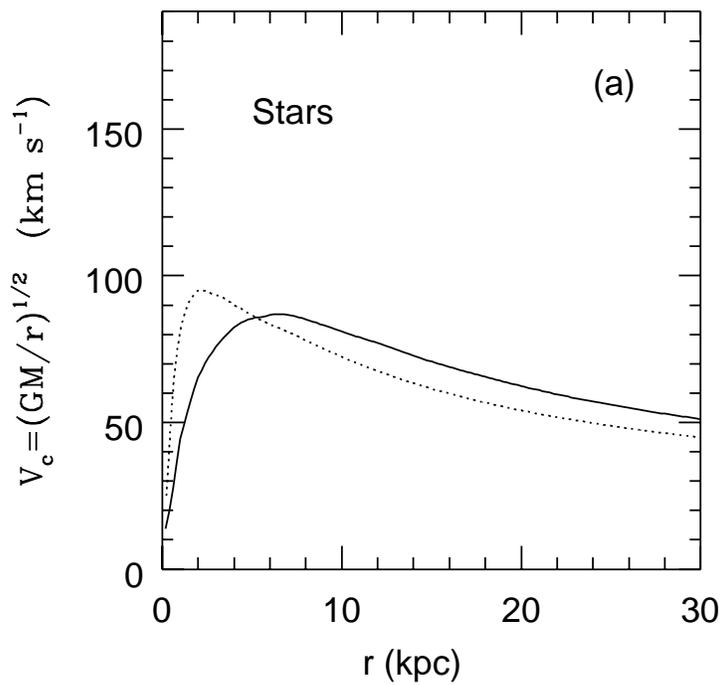
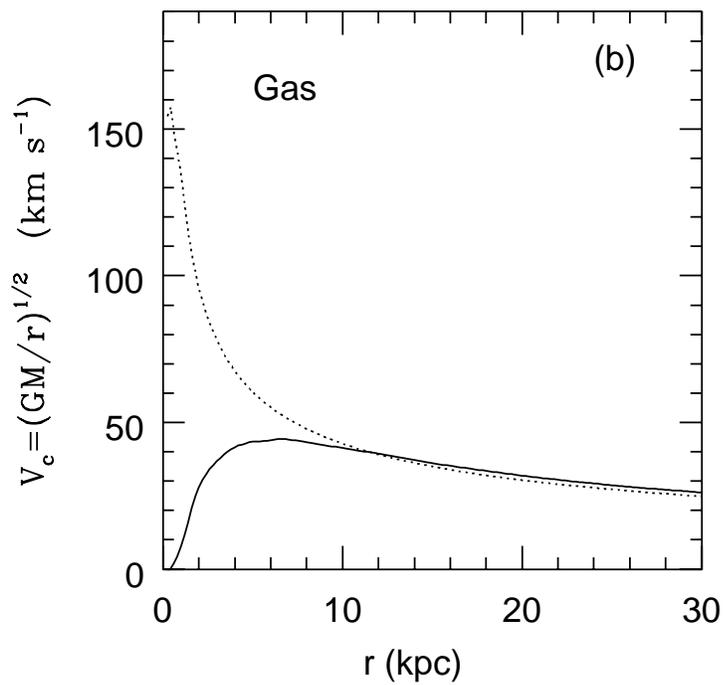
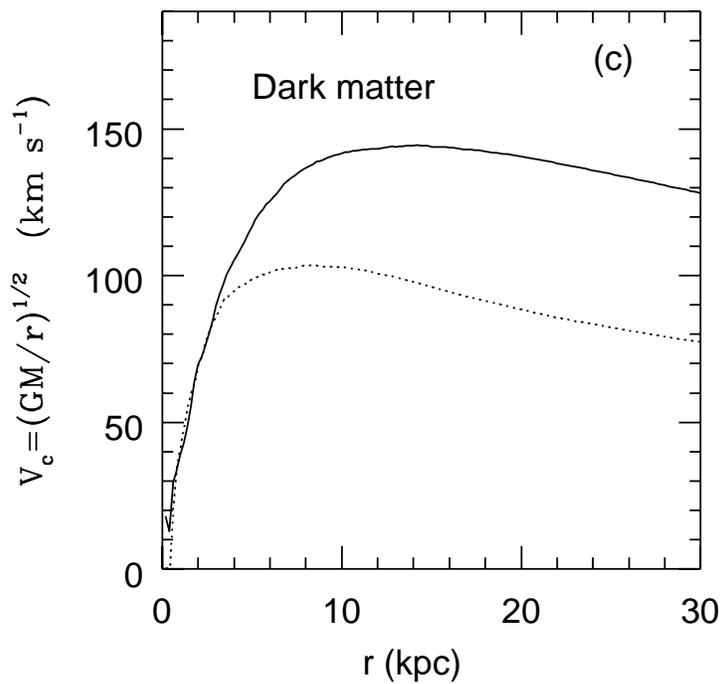
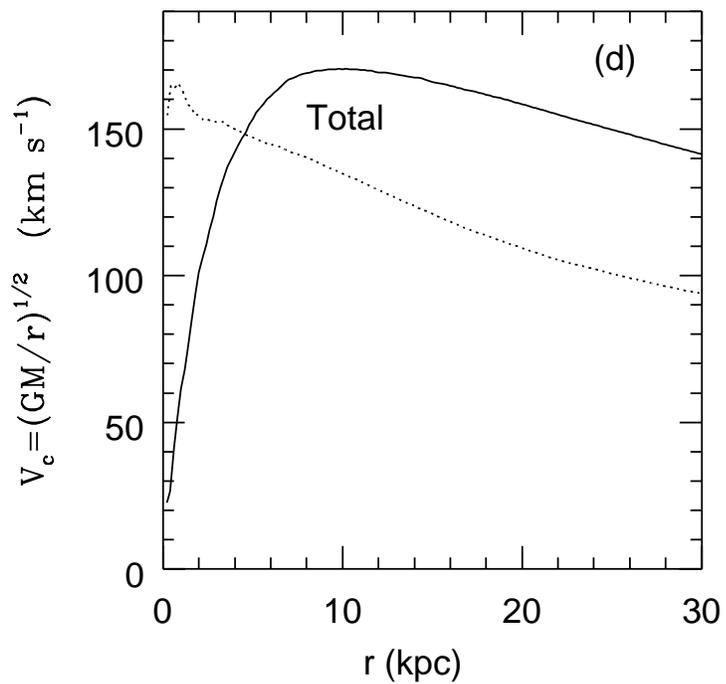
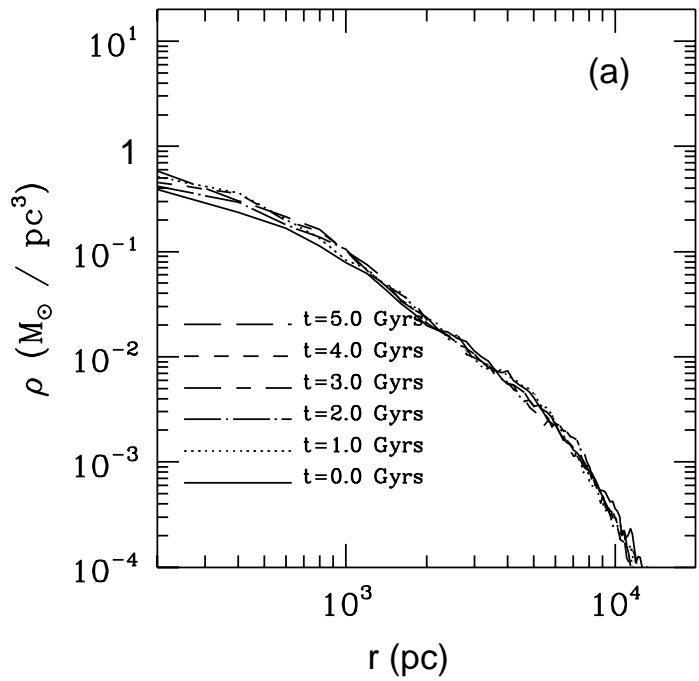
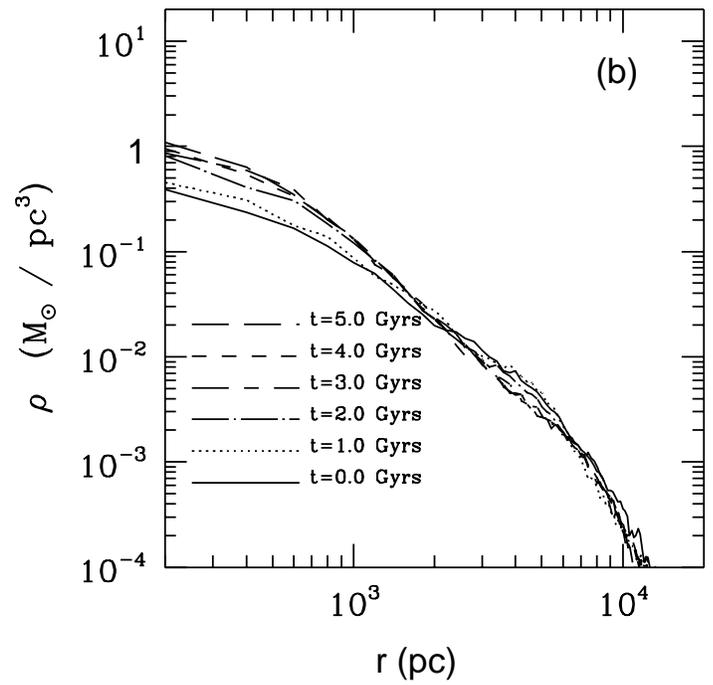
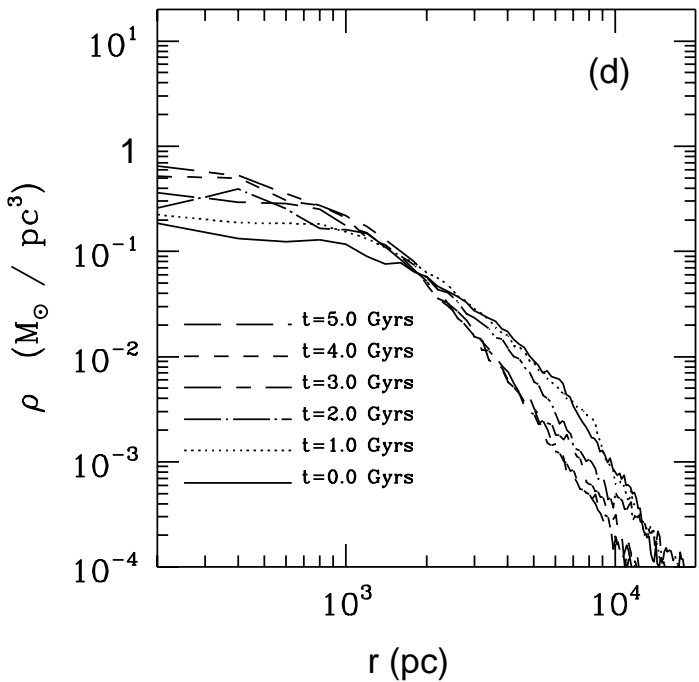
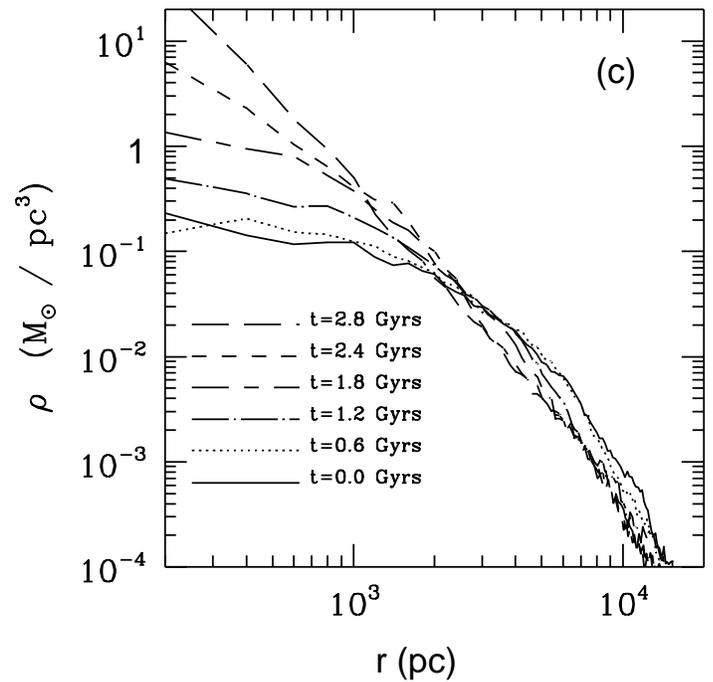
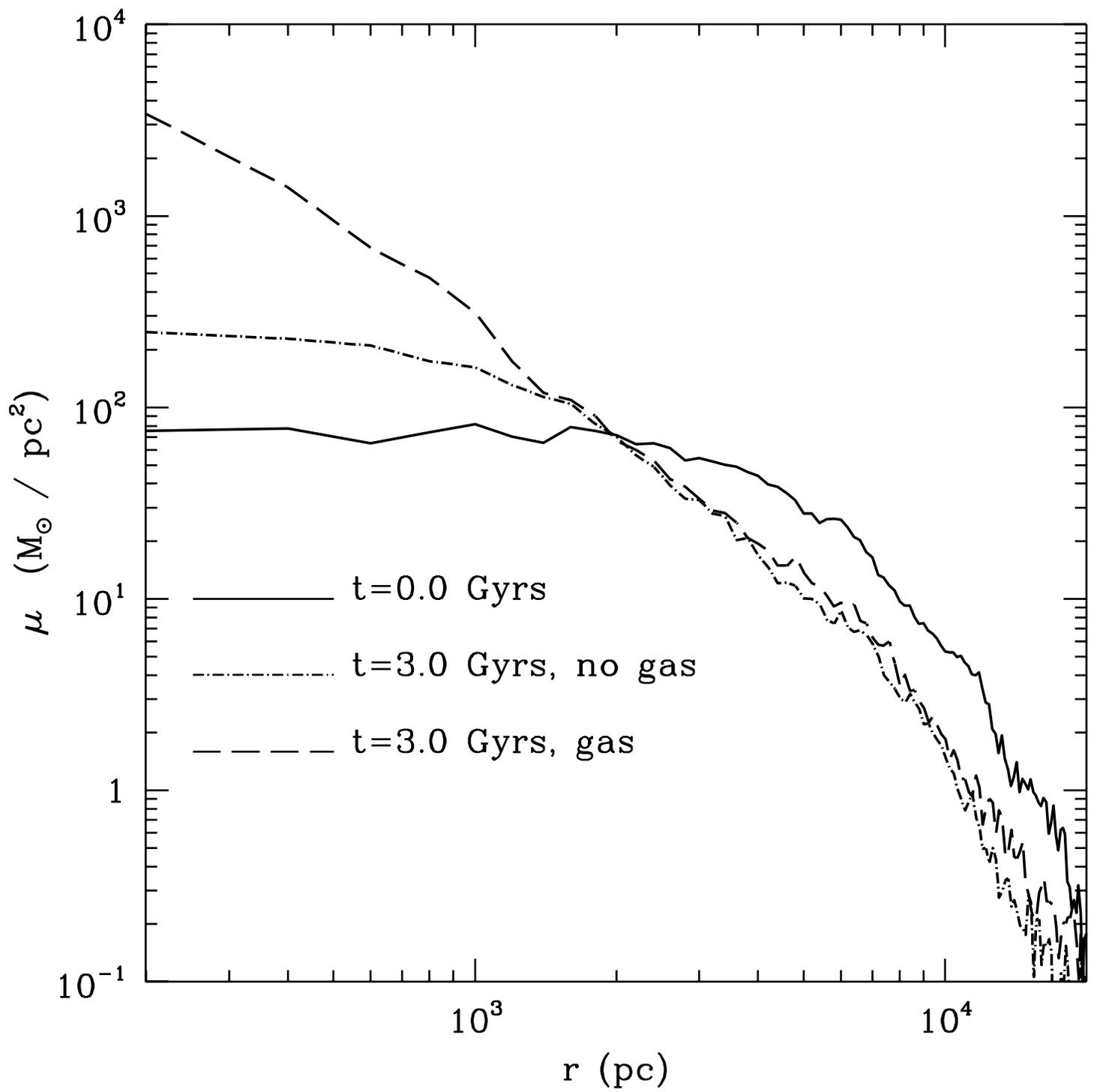
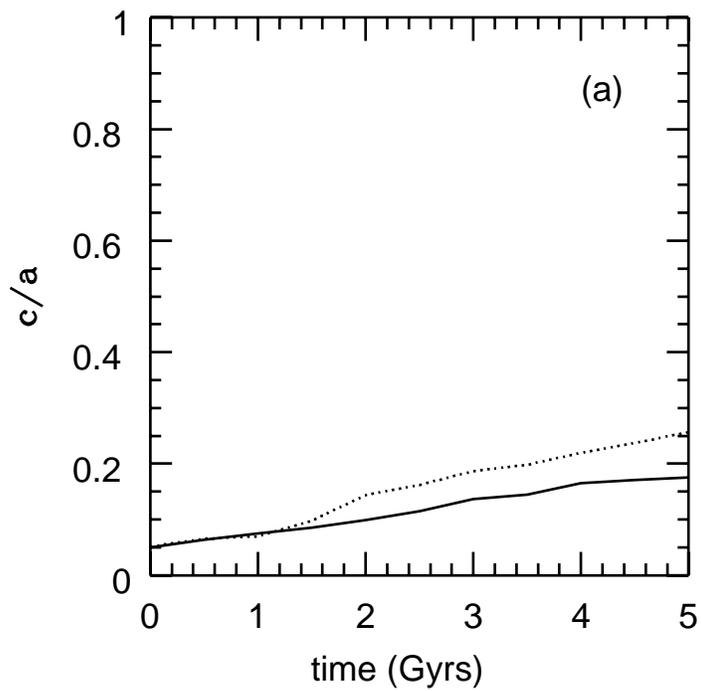
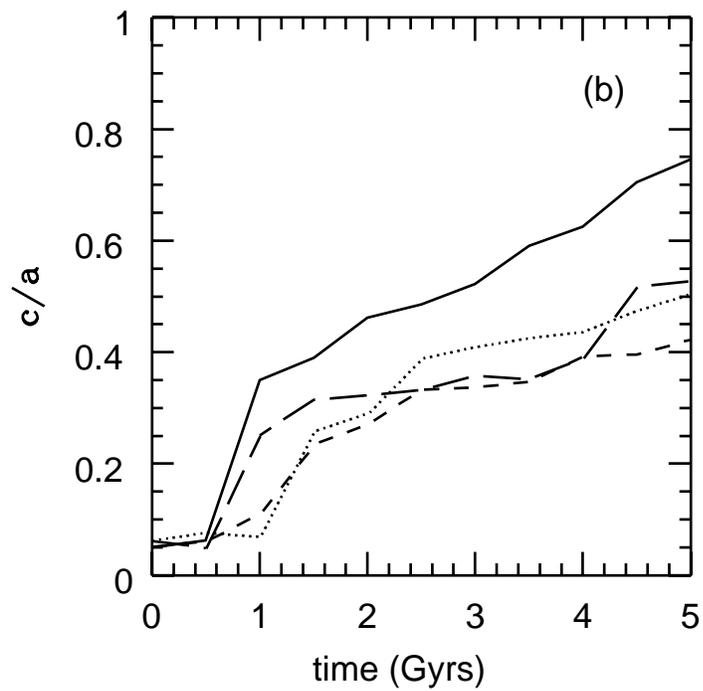
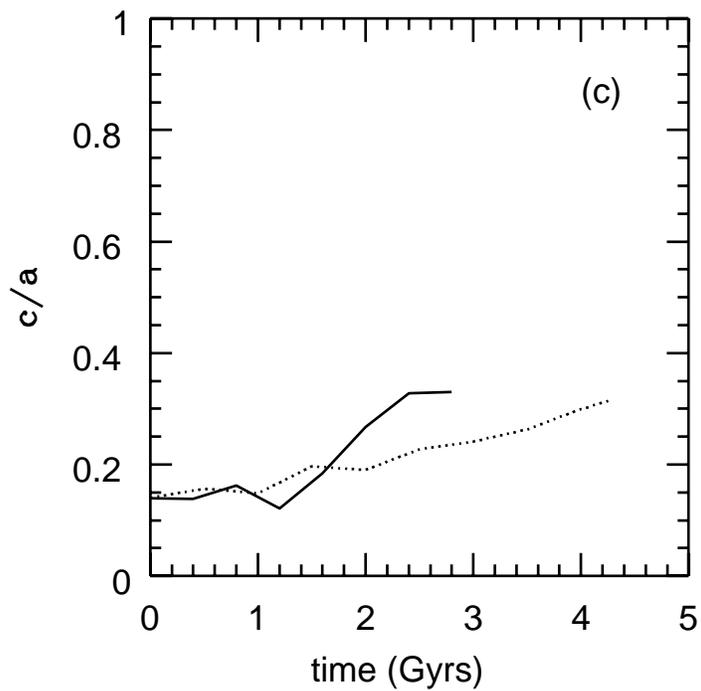
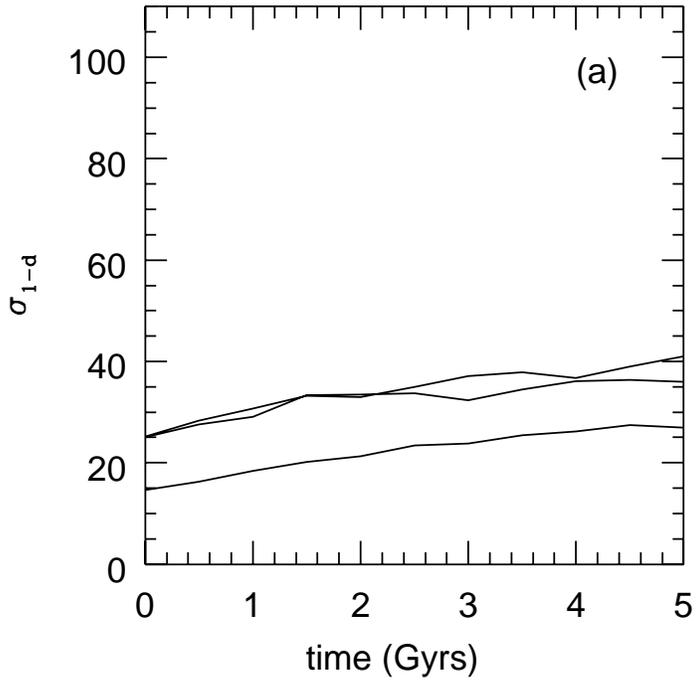
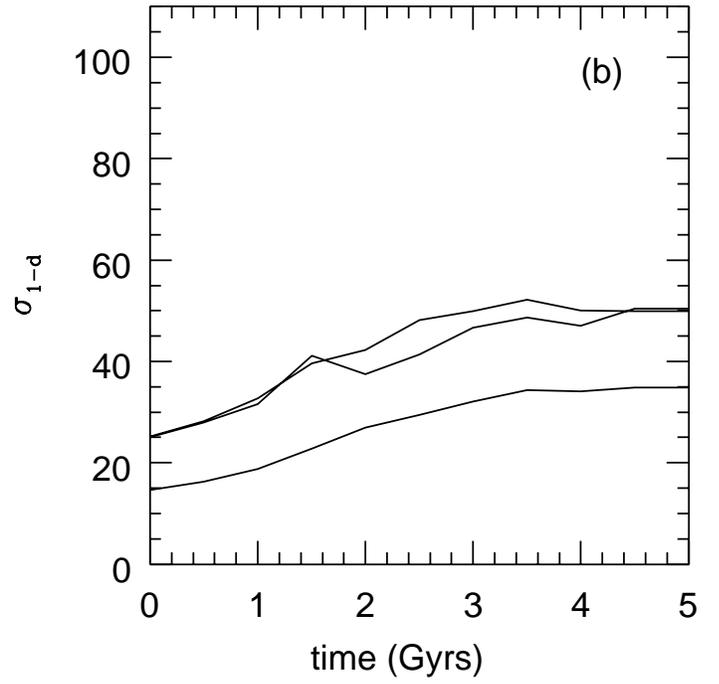
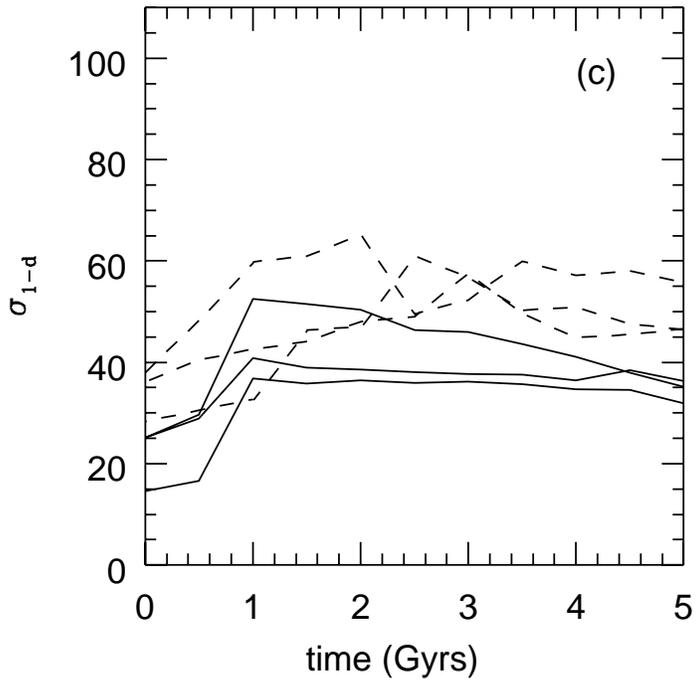
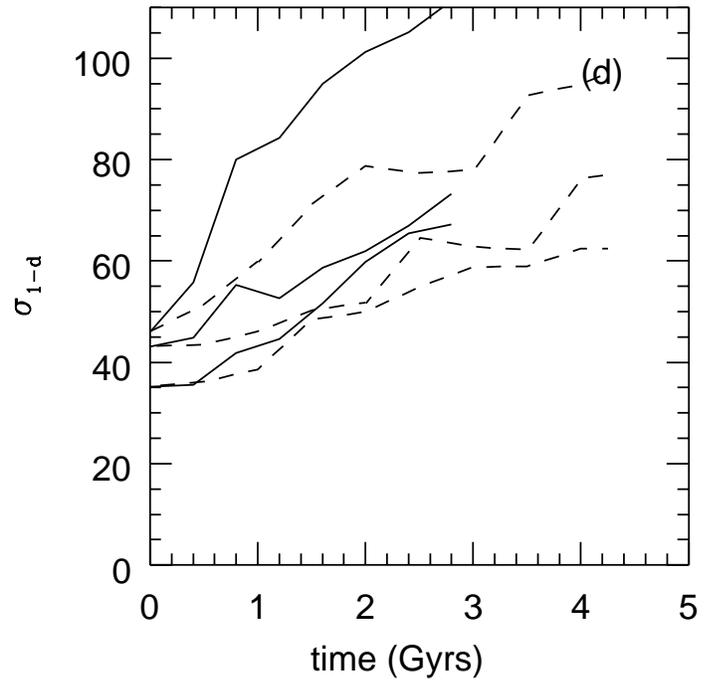
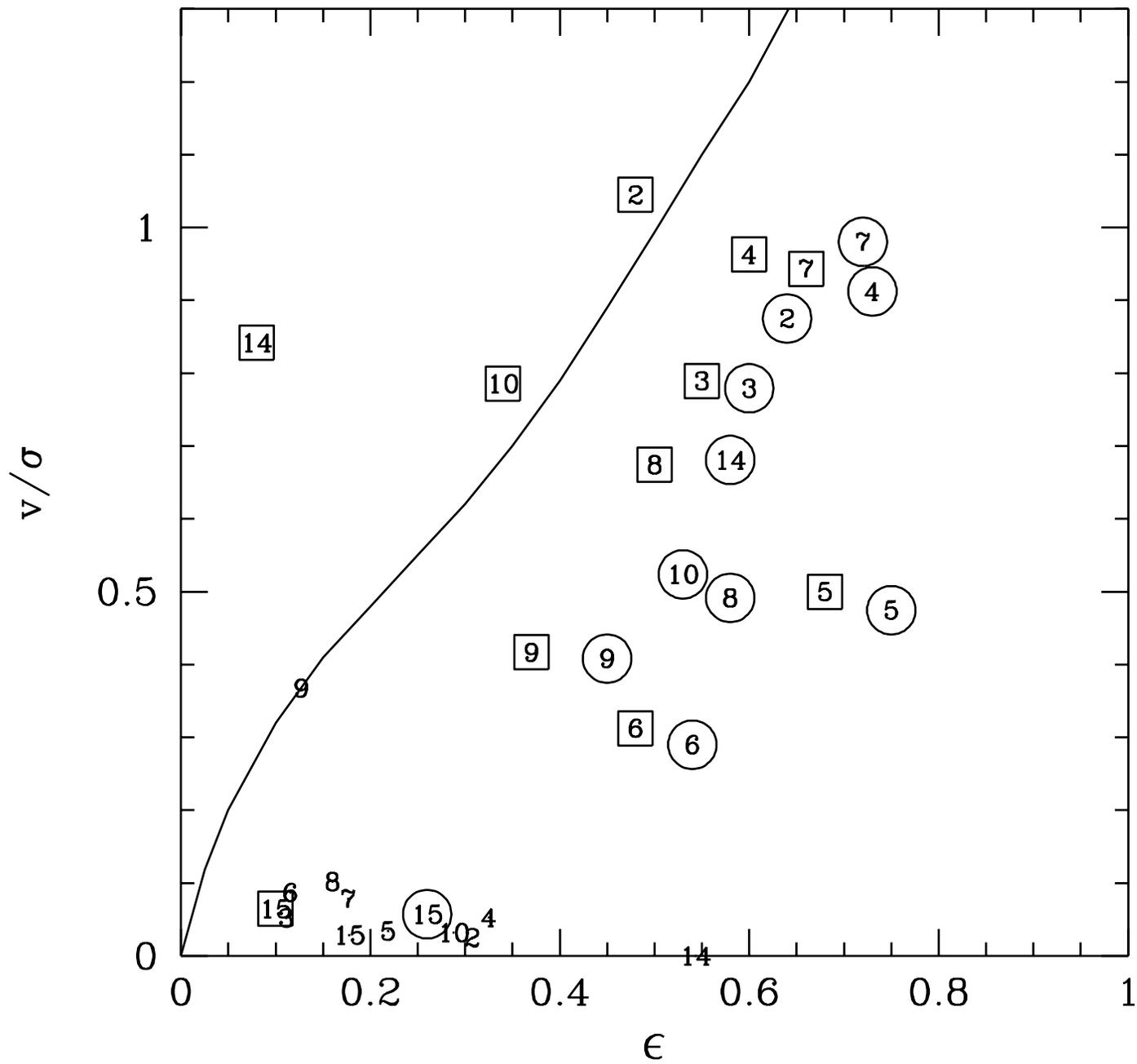
\end{document}